\documentclass[11pt,pdftex]{article}
\usepackage{geometry} 
\usepackage{amsmath}
\usepackage{amssymb}
\usepackage[pdftex]{graphicx}

\usepackage[colorlinks,hyperindex]{hyperref}
\usepackage{thumbpdf}

\usepackage{times}
\usepackage[T1]{fontenc}
\usepackage[latin1]{inputenc}

\newcommand{\TO}{TeO$_2$~}

\begin{document}

\title{Surface-sensitive macrobolometers \\for the identification of external charged particles}
\author{Luca Foggetta, Andrea Giuliani, Claudia Nones,\\  Marisa Pedretti, Samuele Sangiorgio \\ 
\\
Department of Physics and Mathematics, University of Insubria  \\ and INFN-Milano, Via Valleggio 11, 22100 Como, Italy}
\date{}

\maketitle
   
 \begin{abstract}
We report the performance of two prototype \TO macrobolometers, operated at  $\sim$25 mK, able to identify events due to energy deposited at the detector surface. This capability is obtained by thermally coupling thin Ge active layers to the main energy absorber of the bolometer, and is demonstrated by irradiating the detectors with $\alpha$ particles. The temperature variations of the main absorber and of the active layer are measured independently with doped Ge thermistors. These results show clearly that an intrinsic limitation of monolithic low temperature calorimeters, e.g., the impossibility to give information about event position, can be efficiently overcome using composite structures. 

\copyright  \emph{ Copyright (2005) American Institute of Physics.} \footnote{This article may be downloaded for personal use only. Any other use requires prior permission of the author and the American Institute of Physics. This article appeared in \emph{Appl. Phys. Lett. 86, 134106 (2005)} and may be found at \url{http://link.aip.org/link/?apl/86/134106}}
 \end{abstract}

Phonon-mediated detectors (PMD) of atomic and nuclear radiation operated at low temperatures (defined commonly \emph{bolometers}) provide in general better energy resolution, lower energy thresholds and wider material choice than conventional  devices \cite{1}. The best results in terms of energy resolution are obtained when PMDs are operated as perfect calorimeters \cite{2,3,4}. One of the drawback of the calorimetric mode is that only the deposited energy is measured and usually no additional information on the nature of the event can be achieved. For instance, it is not possible to get any space resolution. This limitation is intrinsic to the detection mechanism, since all the deposited energy is degraded and measured in the final form of heat. Some space resolution was demonstrated if out-of-equilibrium phonons are detected \cite{5}, but at the price of a lower energy resolution. In this letter, we show that the dynamics of the heat flow in the detector can be used in order to attain information on the particle impact point, thus overcoming one of the traditional limitation of  low temperature calorimeters. 

The motivation for this work stems from a fundamental physics research carried out in the Gran Sasso underground laboratory (Italy) \cite{6}. In these experiments, TeO2 macrobolometers (with masses up to $\sim$1 kg) are operated at $\sim$10 mK to search for the neutrinoless double beta decay of $^{130}$Te, a rare nuclear process crucial for neutrino physics \cite{7}. The signature for the decay consists of a peak in the energy spectrum of the detector located at 2528 keV. Since the expected decay rate is less than $10^{-2}$ counts/(year mole), the background due to residual radioactive impurities nearby the detector can easily overcome the signal. $\alpha$ and $\beta$ particles emitted close to the surface of the detectors or of the materials surrounding it are particularly dangerous. Since these particles release only a part of their energy in the detector, they contribute with a continuum spectrum which extends down to the region relevant for double beta decay. The safest way to get rid of this effect is to make the detector surface sensitive. In order to develop this capability, we propose a composite detector, consisting of a main \TO bolometer and six (one for each face) thin, large-area, auxiliary Ge or Si bolometers that act as active layers. (These materials are chosen because of their availability, the high achievable purity and the favorable thermal properties.) Each layer will have the same area and shape as the corresponding absorber face and will be separated from it by a sub-millimeter gap, providing a $4\pi$ hermetic coverage from external charged particles. The original solution proposed here is to thermally couple the Ge or the Si layers to the main absorber. On each layer, a thermistor is attached for temperature reading. The origin of the events can be clearly determined by comparing  the amplitude of the pulses from the different detector elements. If an $\alpha$ particle comes from outside the bolometer, it interacts with an active layer releasing there all its energy ("surface'' event). As a consequence, the temperature will rise and there will be a signal on both the layer thermistor and the \TO one. Because of the small heat capacity of the layer, its thermistor signal will be much higher and faster than that of the \TO thermistor. On the other hand, an event inside the \TO crystal ("bulk'' event) will lead to pulses with similar amplitudes and shapes on both thermistors. 

\begin{figure}[tbp]
\centering
 \includegraphics[width=0.3\textwidth]{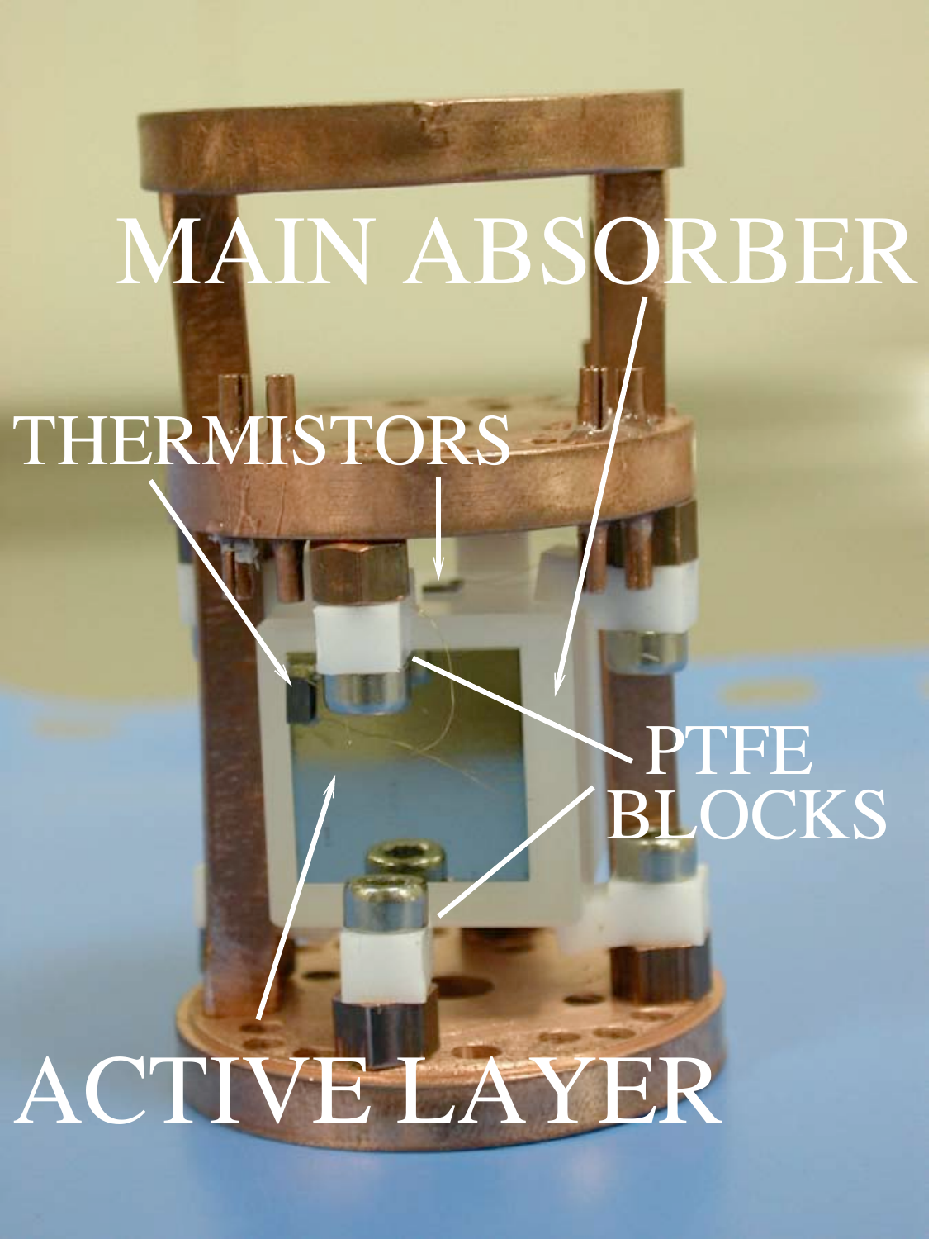}
\caption{This image shows a composite TeO2 bolometer with a Ge active layer (detector A). The NTD thermistor gold wires are not connected to the read-out pads.}
 \label{fig:1}
\end{figure}

In order to test the exposed principles, two prototype detectors (named A and B from now on) were realized. In both of them the main absorber consisted of a \TO single crystal, with a mass of 48 g for detector A and of 12 g for detector B. Detector A absorber was a cube with 2 cm side, while detector B absorber had a rectangular shape with $2\times2\times0.5$ cm dimension. A Ge single crystal with a $1.5\times1.5$ cm surface and 500 $\mu$m thickness was glued at a $2\times2$ cm face by means of four epoxy spots (1 mm diameter and 50 $\mu$m thick). Neutron Transmutation Doped (NTD) Ge thermistors8 were glued at the main absorbers and at the auxiliary Ge bolometers with six epoxy spots (0.5 mm diameter and 50 $\mu$m thickness). The NTD chip size was $3\times1.5\times1$ mm in all cases. The resistance-temperature behavior of the chips \cite{8} is parameterized as $R(T) = R_{0} \exp [ (T_{0} / T )^{1/2}]$,  as predicted by the so called Hopping Conduction Regime with Coulomb gap \cite{9}. The values of $R_{0}$ and $T_{0}$ are respectively 2.65 $\Omega$ and 7.8 K. The coupling of the main absorber to the heat bath was realized by means of 8 (4) PTFE blocks for detector A (B). Detector A is depicted in Fig. \ref{fig:1}. The Ge active layers were exposed to $\alpha$ particles. The source was obtained by a shallow implant of $^{224}$Ra nuclides onto a piece of copper tape, facing the Ge layer. $^{224}$Ra is an $\alpha$ emitter with a half life of 3.66 d in equilibrium with its $\alpha$ and $\beta$ emitting daughters. The main $\alpha$ lines are at 5.68, 6.29, 6.78 and 8.78 MeV. Two weak lines sum up at about 6.06 MeV. 

The detectors were cooled down in two separate runs in a low power dilution refrigerator located in the Cryogenic Laboratory of the Insubria University (Como, Italy). The detector thermistors are DC biased through a voltage supply and a 20 G$\Omega$  room-temperature load resistance. Typical operation temperatures were $T \approx 25$ mK, corresponding to a thermistor resistance $R \approx 100$ M$\Omega$. The voltage across the thermistor is typically $\sim$25 mV. Voltage pulses are read-out by a DC-coupled low noise differential voltage amplifier, followed by a filtering single-ended stage \cite{10}. The front end electronics was located at room temperature. The signals were acquired by a 12 bit transient recorder, collecting 1024 voltage points for each pulse, and registered for off-line analysis. The set-up was not optimized for high energy resolution measurements, since the main purpose of the experiment was to verify and to understand the detector surface sensitivity.

\begin{figure}[tbp]
\centering
 \includegraphics[width=0.65\textwidth]{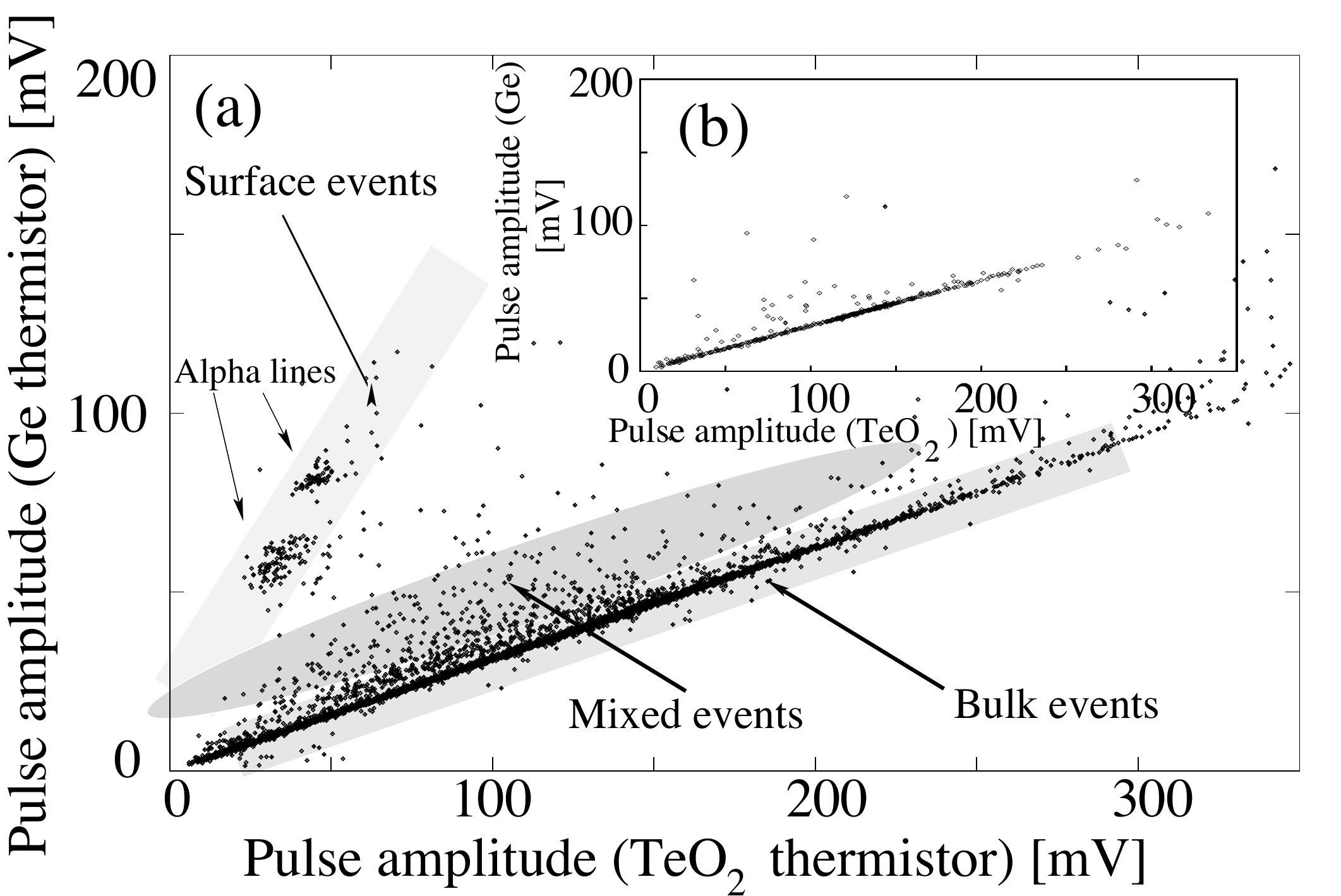}
\caption{(a) Scatter plot showing the relationship between the amplitudes of the pulses acquired in coincidence from the main absorber (X-axis) and from the active layer (Y-axis) for detector A; (b) the same plot after imposing coincidences with pulses from detector C. Not only the surface ? events have been cut, but also "mixed" events due to cosmic muons releasing their energy both in the main absorber and in the active layer.}
 \label{fig:2}
\end{figure}

In the first run, we operated detector A together with a conventional "naked'' bolometer (detector C) having the same features as detector A but no active layer. The two detectors were aligned vertically and the distance between their \TO absorber centers was cm 7. Detector A was irradiated by the source described above. Results in terms of surface/bulk event separation are reported in the scatter plot of Fig. \ref{fig:2} (a), where the clusters due to external $\alpha$ particles and the band due to bulk events are clearly appreciable. Bulk events are due to natural radioactivity and mainly to cosmic muons crossing the \TO absorber. Many points appear in the region between the two bands, in particular immediately above the bulk event band. We make the hypothesis that these points are due to cosmic muons crossing both the \TO absorber and the Ge layer, giving rise to "mixed'' events, with most of the energy deposited in the main absorber but with a small simultaneous energy release in the layer. The ratio of the deposited energy in the two elements is of the order of  (2 cm / 0.05 cm) $\approx$ 40, in favor of the main absorber. This hypothesis can be confirmed by requiring an additional coincidence with detector C. In this fashion, the almost "vertical'' muons crossing both detectors are selected, decreasing dramatically the probability that a muon crosses also the Ge layer, which is placed in the vertical plane. The result of this selection is appreciable in fig. \ref{fig:2} (b), where most of the mixed events above the bulk event band have been cut. This test is very important for our final application. In fact, $\alpha$ particles emitted by the \TO surfaces are also very dangerous for the double beta decay search, since they could escape from the absorber depositing only a fraction of their energy (say $\sim 2.5$ MeV) in it and simulating a double beta decay event. The presence of the active layer however allows to intercept simultaneously the remaining energy, giving rise to a "mixed'' event that will lie in the region between the two bands in the scatter plot, as shown above. In this case, the mixed events will lie close to the surface-event band, since the ratio of deposited energy will range now between (4 - 2.5)/2.5 $\approx$ 0.6 and (9 - 2.5)/2.5 $\approx$ 2.6  (assuming a range $\sim$ 4 - 9 MeV for natural $\alpha$ particle energy), and so much more in favor of the layer with respect to the cosmic muon case. 

\begin{figure}[tbp]
\centering
 \includegraphics[width=0.65\textwidth]{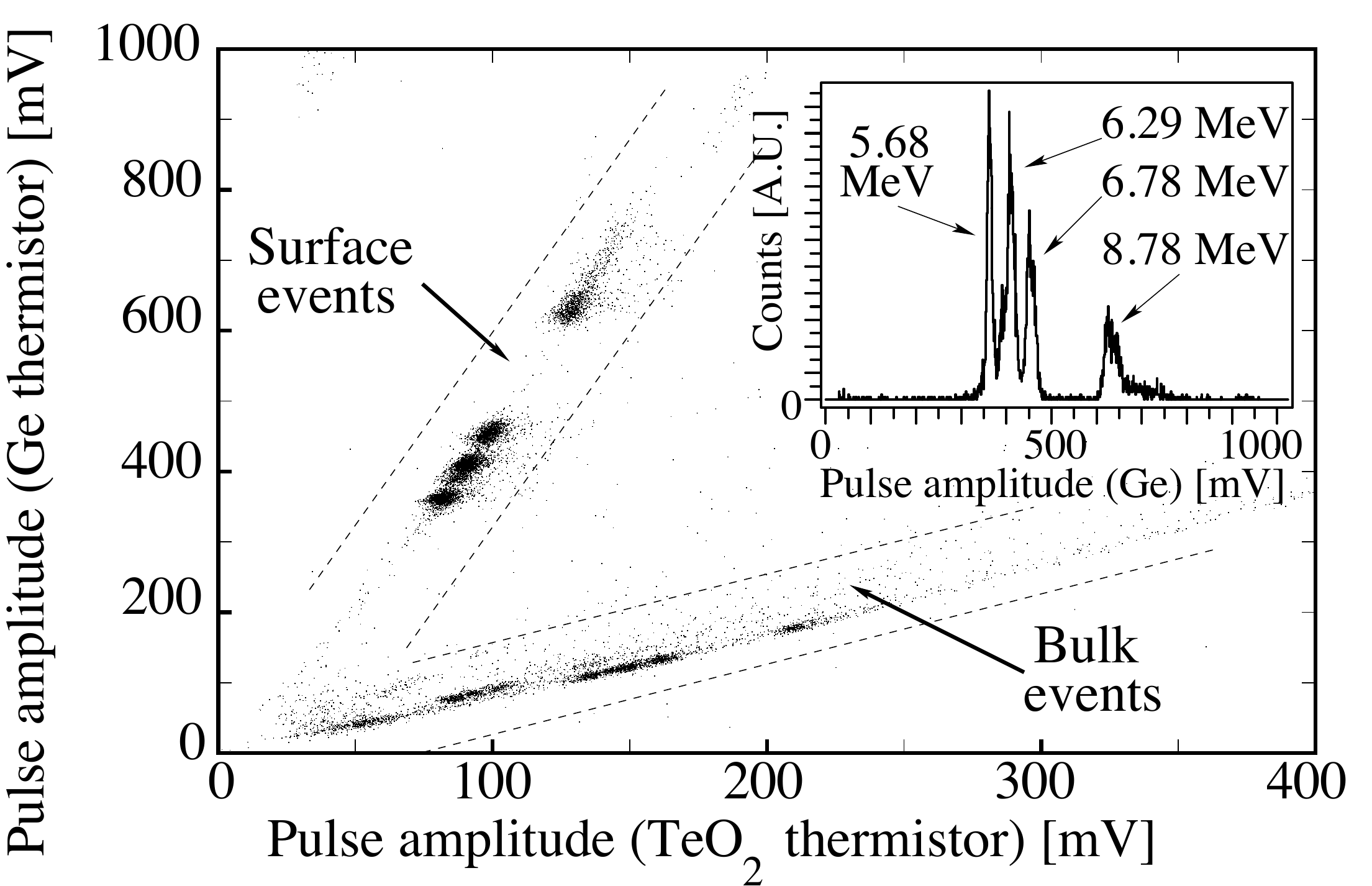}
\caption{Scatter plot showing the relationship between the amplitudes of the pulses acquired in coincidence from the main absorber (X-axis) and from the active layer (Y-axis) for detector B. In the inset, the energy spectrum obtained by selecting surface events is shown. The expected structure of the ? source spectrum emerges clearly, proving that the selection of the surface events is effective.}
 \label{fig:3}
\end{figure}

\begin{figure}[tbp]
\centering
 \includegraphics[width=0.65\textwidth]{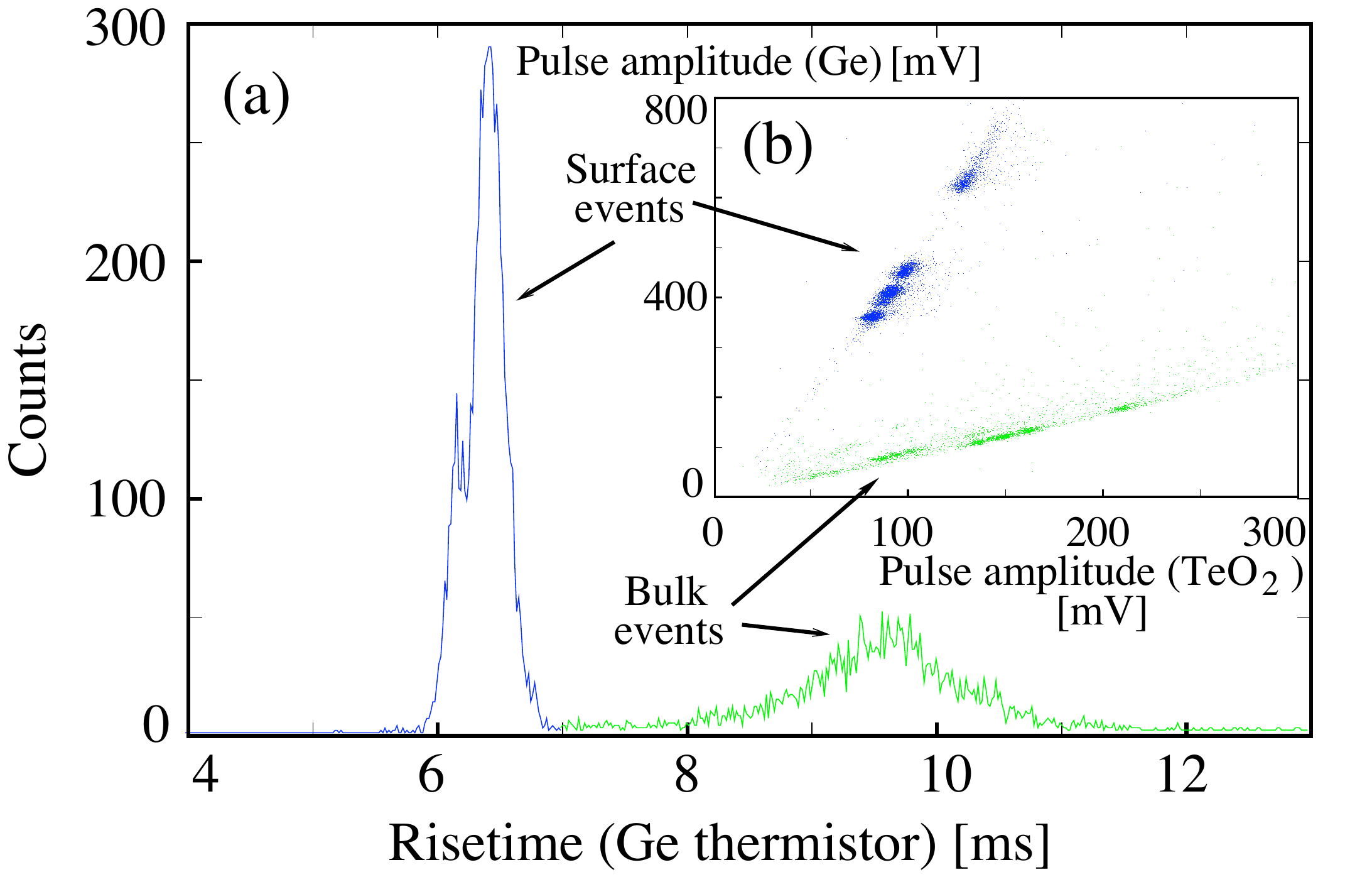}
\caption{(a) Rise time distribution for the pulses developed across the active layer thermistor in detector B. (b) A detail of the scatter plot reported in Fig. \ref{fig:3} where fast active-layer pulses (left Gaussian in (a)) are reported in blue, while slow active-layer pulses (right Gaussian in (a)) in green.}
 \label{fig:4}
\end{figure}

In the second run, we cooled down detector B. No additional detector was present this time. Some improvements were made with respect to the first run: we achieved a better control of the microphonic noise; we increased the source rate and added a second source irradiating directly the main absorber in order to increase the population of the bulk event band in the $\alpha$ region; we enlarged the bandwidth of the acquired signals from DC-12 Hz to DC-120 Hz. This last measure allowed to acquire without electronic integration the fast pulses coming from the layer (rise time $\sim$ 6.5 ms, decay time $\sim$ 40 ms), while it had smaller effects on the slow pulses coming from the main absorber (rise time $\sim$ 9.7 ms, decay time $\sim$ 70 ms). The scatter plot obtained this time is shown in Fig. \ref{fig:3}. The two bands are now much better separated with respect to the first run, and the 4 main $\alpha$ lines of the source are clearly appreciable. A powerful separation between the two event types can be achieved also by rise time selection. In Fig. \ref{fig:4} (a) the rise time distribution is shown for pulses from the layer thermistor. The fast surface events are cleanly separated by the slow bulk events. A selection of the ``fast'' events identifies clearly the surface event band, as shown in Fig. \ref{fig:4} (b).

This experiment demonstrates the power of the technique we have developed to identify a superficial impact point in low temperature calorimeters. In the future, a full coverage device will be realized. The thermistors on the six layers will be acquired in parallel to avoid proliferation of read-out channels. The rise time analysis shows that the scatter plot is redundant to select surface events. It might be convenient then to get rid also of the main absorber thermistor, reducing the read-out channel to one as in the conventional case. On the other hand, a detailed analysis will be performed to check if pulse shape discrimination can allow to identify surface events using signals only from the main absorber thermistor. In this case, the Ge or Si layers would operate as signal-shape modifiers and would not require additional thermistors, leading to a substantial simplification of the device. 

Thanks are due to Oliviero Cremonesi, who developed the programs used for the pulse analysis, to Gianluigi Pessina, who realized the front end electronics, and to Jeff Beeman, who provided us with the NTD thermistors and the Ge layers. The results exposed in this letter are part of a research program funded by the European Commission in the VI Framework Programme, under contract number 50622/RII3-CT-2003-505818, devoted to astroparticle physics and including a specific activity on double beta decay.


\begin{thebibliography}{99}

\bibitem{1} A.~Giuliani, Physica B 280, 501 (2000).
\bibitem{2} A.~Alessandrello, C. Brofferio, C.  Bucci, E.  Coccia, O. Cremonesi, V. Fafone, E.~Fiorini, A.~Giuliani, A. Nucciotti, S. Parmeggiano, M. Pavan, G. Pessina, S. Pirro, E. Previtali, A. Rotilio, M. Vanzini, and L. Zanotti,   Nucl. Instr. Meth. Phys. Res. A 440, 397 (2000).
\bibitem{3} D.A. Wollman, K.D. Irwin, G.C. Hilton, L.L. Dulcie, D.E. Newbury, and J.~M.~Martinis, J. Microscopy 188, 196 (1997).
\bibitem{4} A. Alessandrello, J.W. Beeman, C. Brofferio, C. Bucci, O.~Cremonesi, E.~Fiorini, A.~Giuliani, E.E. Haller, A. Monfardini, A. Nucciotti, M. Pavan, G. Pessina, E. Previtali, and L.  Zanotti, Phys. Rev. Lett. 82, 513 (1999).
\bibitem{5} R. M. Clarke, P. L. Brink, B. Cabrera, P. Colling, M. B. Crisler, A. K. Davies, S.~Eichblatt, R. J. Gaitskell, J. Hellmig, J. M. Martinis, S. W. Nam, T. Saab, and B.~A.~Young, Appl. Phys. Lett. 76, 2958 (2000).
\bibitem{6} C. Arnaboldi, D. R. Artusa, F. T. Avignone, III, M. Balata, I. Bandac, M. Barucci, J. W. Beeman, C. Brofferio, C. Bucci, S. Capelli, F. Capozzi, L. Carbone, S. Cebrian, O. Cremonesi, R. J. Creswick, A. de Waard, H. A. Farach, A. Fascilla, E. Fiorini, G. Frossati, A. Giuliani, P. Gorla, E. E. Haller, R. J. McDonald, A. Morales, E.~B.~Norman, A. Nucciotti, E. Olivieri, E. Palmieri, E. Pasca, M. Pavan, M. Pedretti, G.~Pessina, S. Pirro, C. Pobes, E. Previtali, M. Pyle, L. Risegari, C. Rosenfeld, S. Sangiorgio, M. Sisti, A. R. Smith, L. Torres, and G. Ventura, Phys. Lett. B 584, 260 (2004).
\bibitem{7} S.R. Elliot and P. Vogel, Ann. Rev. Nucl. Part. Sci. 52, 115 (2002).
\bibitem{8} E. Haller, J. Appl. Phys. 77, 2857 (1995).
\bibitem{9} B.I. Shklovskii and A. L. Efros, Sov. Phys. - JETP 33, 468 (1971).
\bibitem{10} A. Alessandrello, C. Brofferio, C. Bucci, D. V.  Camin, O.  Cremonesi, A. Giuliani, A.~Nucciotti, M. Pavan, G. Pessina, E. Previtali, and G. Sablich, IEEE Trans. Nucl. Sci. 44, 416 (1997).
\end{thebibliography}
\end{document}